\documentclass{aa}
\usepackage{graphicx}
\usepackage{amsfonts,amssymb,latexsym}

\def\lambdabar{\protect\@lambdabar}
\def\@lambdabar{%
\relax
\bgroup
\def\@tempa{\hbox{\raise.73\ht0
\hbox to0pt{\kern.25\wd0\vrule width.5\wd0
height.1pt depth.1pt\hss}\box0}}%
\mathchoice{\setbox0\hbox{$\displaystyle\lambda$}\@tempa}%
{\setbox0\hbox{$\textstyle\lambda$}\@tempa}%
{\setbox0\hbox{$\scriptstyle\lambda$}\@tempa}%
{\setbox0\hbox{$\scriptscriptstyle\lambda$}\@tempa}%
\egroup
}
\def\chem#1#2{$\rm{}^{#1}\kern-0.8pt#2$}
\def\reac#1#2#3#4#5#6{$\rm\,{}^{#1}\kern-0.8pt{#2}\,({#3}\,,{#4})\,
{}^{#5}\kern-0.8pt{#6}\,$}
\def\gsimeq{\,\,\raise0.14em\hbox{$>$}\kern-0.76em\lower0.28em\hbox
{$\sim$}\,\,}
\def\lsimeq{\,\,\raise0.14em\hbox{$<$}\kern-0.76em\lower0.28em\hbox
{$\sim$}\,\,}
\def\be{\begin{equation}} 
\def\ee{\end{equation}}
\def\beqy{\begin{eqnarray}}
\def\eeqy{\end{eqnarray}}
\def\bmlet{\begin{mathletters}}
\def\emlet{\end{mathletters}}

\def\A&A#1#2#3{ {\it Astron. Astrophys.} {\bf #2}, #3 (#1)}

\begin{document}

%
\title{BRUSLIB and NETGEN: the Brussels nuclear reaction rate library and nuclear
network generator for astrophysics}

   \author{M. Aikawa, M. Arnould, S. Goriely, A. Jorissen and K. Takahashi}

   \offprints{M. Arnould}

   \institute{Institut d'Astronomie et d'Astrophysique, Universit\'e
              Libre de Bruxelles, CP 226, B-1050 Brussels, Belgium.\\
              Email:  marnould@astro.ulb.ac.be
                     }

    \date{Received --; accepted --}

\abstract{Nuclear reaction rates are quantities of fundamental importance in astrophysics. Substantial
efforts have been devoted in the last decades to measure or calculate them. The present paper
presents for the first time a detailed description of the Brussels nuclear reaction rate library
BRUSLIB and of the nuclear network generator NETGEN in a journal and in a format so as to make these
nuclear data packages easily accessible to astrophysicists for a large variety of applications. BRUSLIB
is made of two parts. The first one contains the 1999 NACRE compilation based on experimental data for
86 reactions with (mainly) stable targets up to Si. BRUSLIB provides an electronic link to the
published, as well as to a large body of unpublished, NACRE data containing adopted rates, as well as
lower and upper limits to these rates. The second part of BRUSLIB concerns nuclear reaction
rate predictions that have to complement the experimentally-based rates for use in a large variety of
astrophysics modellings. An electronic access is provided to tables of thousand of rates  
calculated within a statistical Hauser-Feshbach approximation, which limits the reliability of
the rates to reactions producing compound nuclei with a high enough level density. These calculations
make use of {\it global} and {\it coherent} {\it microscopic} nuclear models for the quantities
entering the rate calculations. The use of such models is utterly important, and makes the BRUSLIB rate
library unique.
 A description of the Nuclear Network Generator NETGEN that complements the BRUSLIB
package is also presented. NETGEN is a tool to generate nuclear reaction rates for temperature grids
specified by the user. The information it provides can be used for a large variety of applications,
including Big Bang nucleosynthesis, the energy generation and nucleosynthesis associated with the
non-explosive and explosive hydrogen to silicon burning stages, or the synthesis of the heavy nuclides
through the s-,
$\alpha$- and r-, rp- or p-processes.   
}

\titlerunning{The Brussels nuclear reaction rate library}
\authorrunning{M. Aikawa et al.}

\maketitle
%

\section{Introduction}
\label{intro}
 
Since around the nineteen fifties, astrophysics has advanced at a remarkable pace, and has
achieved an impressive record of success. One of the factors contributing to these rapid
developments is undoubtedly a series of spectacular breakthroughs in nuclear   
astrophysics, which embodies the special interplay between nuclear physics and
astrophysics.  

The close relationship between these two major scientific disciplines comes about because of
the clear demonstration that the Universe is pervaded with nuclear physics imprints at
all scales. Starting with the Big Bang nucleosynthesis episode, the structure, evolution and
composition of a large variety of cosmic objects, including the Solar System and its
various constituants (down to meteoritic grains), bear strong imprints of the properties of
atomic nuclei, as well as of their interactions.  Therefore, careful and dedicated
experimental and theoretical studies of a large variety of nuclear processes are
indispensable tools for the modeling of the ultra-macroscopic systems  astrophysics has to deal
with.

Over the years, an impressive body of nuclear data of astrophysics interest have been
obtained through laboratory efforts, complemented with theoretical developments. The latter
are indispensable as nuclear experiments conducted today or expected to be performed in any
foreseeable future can and will cover only a minute fraction of the astrophysics needs. This
is because an extremely large variety of highly unstable `exotic' nuclei that, for
long, will not be produced and studied in the laboratory are expected to be involved in the
modelling of many astrophysics processes and events. Many of the basic properties of these
nuclei are to be known for this purpose, and so are their interactions, in particular those with
nucleons or $\alpha$-particles. Even when laboratory-studied nuclei are considered, theory
has very often to be called for assistance. In many respects, laboratory conditions are
indeed very different from stellar ones, which are highly versatile and are often
charaterized by high temperatures and/or densities that are out of reach of laboratory
simulations. In addition, nuclear reactions between charged particles inside 
 non-exploding stars take
place in an energy regime that is in all but a few exceptional cases out of reach of direct 
experimental scrutiny. Indirect methods bring some complement of information, but clearly do not cover
all the needs. In explosive situations, the energies of astrophysical interest are higher, and the
cross sections are correspondingly larger. However, in such events, many reactions
involve unstable targets, so that the fraction of those reactions of potential
interest for which experimental reaction data are lacking is even larger.

The experimental and theoretical nuclear physics achievement mentioned above in fact let
nuclear astrophysics face a new and difficult challenge. The rapidly growing
wealth of nuclear data becomes, ironically, less and less easily accessible to the astrophysics 
community. Mastering this volume of information and making it available in an accurate and
usable form for incorporation into astrophysics models become urgent goals of prime
necessity. The establishment of the required level in the  privileged communication between
nuclear physicists and astrophysicists necessarily requires the build-up of well documented and
evaluated sets of experimental data or theoretical predictions of astrophysical relevance.
Some years ago, the Institut d'Astronomie et d'Astrophysique of the
Universit\'e Libre de Bruxelles has decided to tackle this challenge. This work has grown
into the BRUSsels nuclear LIBrary for astrophysics applications, referred to as BRUSLIB, and
into a nuclear NETwork GENerator called NETGEN.

One goal of this paper is to \emph{present for the first time a detailed description of
BRUSLIB and of NETGEN in a journal and in a format such that 
these nuclear packages become easily accessible to astrophysicists} for a large variety of
applications. In particular, the BRUSLIB reaction rates
presented in a tabular form that is well suited to astrophysics needs are available
electronically at the address {\it http://www-astro.ulb.ac.be}.   
 
BRUSLIB is composed of two main
parts. The first one concerns the Nuclear Astrophysics Compilation of REaction rates referred
to as NACRE.  This compilation provides the rates of 86 thermonuclear reactions of
astrophysical relevance based on an in-depth analysis of {\it experimental data}.
Its description is presented in Sect. \ref{NACRE}. The second part describes {\it theoretical
evaluations} of a collection of about 100000 rates for thermonuclear  reactions induced by nucleons or $\alpha$-particles, as well as
photo-induced reactions not included in NACRE, including nuclei with $8\le Z\le 110$
lying between the proton and the neutron drip lines (Sect.~\ref{th-rates}). The rate
calculations are based on the statistical Hauser-Feshbach (hereafter HF) model. They
require the knowledge of a substantial amount of data concerning basic properties of the
nuclei and of their interactions. The predictions of these properties rely on the use of
{\it global} and {\it coherent} {\it microscopic} nuclear models. These utterly
important model characteristics make the BRUSLIB rate library unique. A confrontation
between selected experimental data and the predictions used in BRUSLIB is provided in
Sect.~\ref{th-exp}. 

A second goal is to describe NETGEN (Sect.~\ref{NETGEN}), a package for
constructing nuclear reaction networks on grounds of the nuclear physics input from BRUSLIB
and, when necessary, from other sources.  

It has to be noted that no information is provided by the present release of BRUSLIB on the rates
of non-statistical (`direct') or of non-thermonuclear (`spallation') reactions that can
develop in low-temperature and low-density astrophysical media, like the interstellar or
circumstellar medium.

\section{The experimentally-based NACRE evaluation and compilation of nuclear reaction rates}
\label{NACRE}

The necessity of placing relevant nuclear data of high quality at the disposal of the
astrophysics community has been the driving motivation for the construction of  a compilation aimed at superseding the work of Fowler and collaborators (see Caughlan
\& Fowler 1988, hereafter CF88, and references therein to former compilations). The goal
was not just to update the CF88 rates with newly available experimental data, but also to modify quite
deeply different aspects of the format of the CF88 compilation. At this point, slightly more
than half of the CF88 rates have been re-compiled on grounds of a careful evaluation  of
{\it experimental data available by 15 June 1998} , and the results of
the work make the so-called NACRE (Nuclear Astrophysics Compilation of REactions) compilation (Angulo
et al. 1999). The  reactions analyzed up to now are listed in Table~1. They comprise an ensemble of 86
charged particle induced reactions on {\it stable} targets up to Si involved in Big Bang
nucleosynthesis and in the non-explosive H- and He-burning modes, complemented with a restricted number
of reactions of special astrophysical significance on the unstable \chem{7}{Be}, \chem{22}{Na} and
\chem{26}{Al} nuclides. An updated and enlarged version of NACRE is currently in preparation.
 
\begin{table*} 
\caption{List of the NACRE compiled reactions. References to compilations, new data or evaluations
concerning some of these reactions that have appeared after the completion of NACRE are provided. A +
sign indicates that only the most recent reference to a given reaction is provided. Reaction rates are given 
in the corresponding reference $(a)$ in tabular form, $(b)$ as analytic formulae, or $(c)$ as a figure.} 
\vskip 0.5cm

\begin{tabular}{ll l l l l l l l l l l l}
 \hline
Reaction & Rate & Other refs. & Reaction & Rate & Other refs. \\
\hline

$^{1}$H(p,$\nu$e$^{+}$)$^{2}$H  	&	  	&	  	&	$^{15}$N(p,$\alpha$)$^{12}$C  	&	  	&	  	\\
$^{2}$H(p,$\gamma$)$^{3}$He  	&	 [D]$^{a}$,[N]$^{a}$ 	&	  	&	$^{15}$N($\alpha,\gamma$)$^{19}$F  	&	  	&	 [WI02]+ 	\\
$^{2}$H(d,$\gamma)^{4}$He  	&	  	&	 [SA04] 	&	$^{16}$O(p,$\gamma$)$^{17}$F  	&	  	&	  	\\
$^{2}$H(d,n)$^{3}$He  	&	 [D]$^{a}$ 	&	  	&	$^{16}$O($\alpha$,$\gamma$)$^{20}$Ne  	&	  	&	  	\\
$^{2}$H(d,p)$^{3}$H  	&	 [D]$^{a}$ 	&	  	&	$^{17}$O(p,$\gamma$)$^{18}$F  	&	  	&	 [FI05]+ 	\\
$^{2}$H($\alpha,\gamma$)$^{6}$Li  	&	  	&	  	&	$^{17}$O(p,$\alpha$)$^{14}$N  	&	  	&	 [FO04] 	\\
$^{3}$H(d,n)$^{4}$He  	&	 [D]$^{a}$ 	&	  	&	$^{17}$O($\alpha$,n)$^{20}$Ne  	&	  	&	  	\\
$^{3}$H($\alpha,\gamma$)$^{7}$Li  	&	 [D]$^{a}$ 	&	  	&	$^{18}$O(p,$\gamma$)$^{19}$F  	&	  	&	  	\\
$^{3}$He($^{3}$He,2p)$^{4}$He  	&	  	&	 [KU04]+ 	&	$^{18}$O(p,$\alpha$)$^{15}$N  	&	  	&	  	\\
$^{3}$He($\alpha$,$\gamma$)$^{7}$Be  	&	 [D]$^{a}$ 	&	 [NA04] 	&	$^{18}$O($\alpha$,$\gamma$)$^{22}$Ne  	&	 [DA03]$^{a}$ 	&	  	\\
$^{4}$He($\alpha$n,$\gamma$)$^{9}$Be  	&	 [SU02]$^{a}$+ 	&	  	&	$^{18}$O($\alpha$,n)$^{21}$Ne  	&	  	&	  	\\
$^{4}$He($\alpha\alpha,\gamma)^{12}$C  	&	 [FY05]$^{c}$ 	&	  	&	$^{19}$F(p,$\gamma)^{20}$Ne  	&	  	&	  	\\
$^{6}$Li(p,$\gamma$)$^{7}$Be  	&	  	&	 [PR04] 	&	$^{19}$F(p,n)$^{19}$Ne  	&	  	&	  	\\
$^{6}$Li(p,$\alpha$)$^{3}$He  	&	  	&	 [TU03] 	&	$^{19}$F(p,$\alpha)^{16}$O  	&	 [SP00]$^{c}$ 	&	  	\\
$^{7}$Li(p,$\gamma$)$^{8}$Be  	&	 [N]$^{a}$ 	&	  	&	$^{20}$Ne(p,$\gamma)^{21}$Na  	&	 [I]$^{a}$ 	&	  	\\
$^{7}$Li(p,$\alpha$)$^{4}$He  	&	 [D]$^{a}$ 	&	  	&	$^{20}$Ne(p,$\alpha$)$^{17}$F  	&	  	&	  	\\
$^{7}$Li($\alpha$,$\gamma$)$^{11}$B  	&	  	&	 [GY04] 	&	$^{20}$Ne$(\alpha,\gamma)^{24}$Mg  	&	  	&	  	\\
$^{7}$Li($\alpha$,n)$^{10}$B  	&	  	&	  	&	$^{21}$Ne(p,$\gamma$)$^{22}$Na  	&	 [I]$^{a}$ 	&	  	\\
$^{7}$Be(p,$\gamma$)$^{8}$B  	&	  	&	 [CY04]+ 	&	$^{21}$Ne($\alpha$,n)$^{24}$Mg  	&	  	&	  	\\
$^{7}$Be($\alpha,\gamma)^{11}$C  	&	  	&	  	&	$^{22}$Ne(p,$\gamma$)$^{23}$Na  	&	 [HA02]$^{a}$,[I]$^{a}$ 	&	  	\\
$^{9}$Be(p,$\gamma$)$^{10}$B  	&	 [N]$^{a}$ 	&	  	&	$^{22}$Ne($\alpha$,$\gamma$)$^{26}$Mg  	&	  	&	  	\\
$^{9}$Be(p,n)$^{9}$B  	&	  	&	  	&	$^{22}$Ne($\alpha$,n)$^{25}$Mg  	&	 [JA01]$^{a}$ 	&	  	\\
$^{9}$Be(p,d)$^{8}$Be  	&	  	&	 [BR98] 	&	$^{22}$Na(p,$\gamma$)$^{23}$Mg  	&	 [JE04]$^{c}$,[I]$^{a}$ 	&	  	\\
$^{9}$Be(p,$\alpha)^{6}$Li  	&	  	&	 [BR98] 	&	$^{23}$Na(p,$\gamma)^{24}$Mg  	&	 [HA04]$^{a}$,[I]$^{a}$ 	&	 [RO04] 	\\
$^{9}$Be($\alpha$,n)$^{12}$C  	&	  	&	  	&	$^{23}$Na(p,n)$^{23}$Mg  	&	  	&	  	\\
$^{10}$B(p,$\gamma)^{11}$C  	&	 [TO03]$^{b}$ 	&	  	&	$^{23}$Na(p,$\alpha$)$^{20}$Ne  	&	 [HA04]$^{a}$,[I]$^{a}$ 	&	 [RO04] 	\\
$^{10}$B(p,$\alpha$)$^{7}$Be  	&	  	&	  	&	$^{23}$Na($\alpha$,n)$^{26}$Al$^{\rm g}$  	&	  	&	  	\\
$^{11}$B(p,$\gamma)^{12}$C  	&	 [N]$^{a}$ 	&	  	&	$^{23}$Na($\alpha$,n)$^{26}$Al$^{\rm m}$  	&	  	&	  	\\
$^{11}$B(p,n)$^{11}$C  	&	  	&	  	&	$^{23}$Na($\alpha$,n)$^{26}$Al$^{\rm t}$  	&	  	&	  	\\
$^{11}$B(p,$\alpha$)$^{8}$Be  	&	  	&	 [SP04] 	&	$^{24}$Mg(p,$\gamma$)$^{25}$Al  	&	 [I]$^{a}$ 	&	  	\\
$^{12}$C(p,$\gamma$)$^{13}$N  	&	  	&	  	&	$^{24}$Mg(p,$\alpha$)$^{21}$Na  	&	  	&	  	\\
$^{12}$C($\alpha$,$\gamma$)$^{16}$O  	&	 [KU02]$^{a}$ 	&	 [TI02] 	&	$^{25}$Mg(p,$\gamma$)$^{26}$Al$^{\rm g}$  	&	 [I]$^{a}$ 	&	  	\\
$^{13}$C(p,$\gamma)^{14}$N  	&	 [MU03a]$^{a}$ 	&	  	&	$^{25}$Mg(p,$\gamma$)$^{26}$Al$^{\rm m}$  	&	 [I]$^{a}$ 	&	  	\\
$^{13}$C(p,n)$^{13}$N  	&	  	&	  	&	$^{25}$Mg(p,$\gamma$)$^{26}$Al$^{\rm t}$  	&	 [I]$^{a}$ 	&	  	\\
$^{13}$C($\alpha$,n)$^{16}$O  	&	 [KU03]$^{b}$ 	&	  	&	$^{25}$Mg($\alpha$,n)$^{28}$Si  	&	  	&	  	\\
$^{13}$N(p,$\gamma)^{14}$O  	&	 [TA04]$^{a}$ 	&	  	&	$^{26}$Mg(p,$\gamma$)$^{27}$Al  	&	 [I]$^{a}$ 	&	  	\\
$^{14}$N(p,$\gamma$)$^{15}$O  	&	 [MU03b]$^{a}$,[AN01]$^{a}$ 	&	 [RU05]+ 	&	$^{26}$Mg($\alpha$,n)$^{29}$Si  	&	  	&	  	\\
$^{14}$N(p,n)$^{14}$O  	&	  	&	  	&	$^{26}$Al$^{\rm gs}$(p,$\gamma$)$^{27}$Si  	&	 [I]$^{a}$ 	&	  	\\
$^{14}$N(p,$\alpha$)$^{11}$C  	&	  	&	  	&	$^{26}$Al$^{\rm ms}$(p,$\gamma$)$^{27}$Si  	&	 [I]$^{a}$ 	&	  	\\
$^{14}$N($\alpha$,$\gamma$)$^{18}$F  	&	 [G\"O00]$^{a}$ 	&	  	&	$^{27}$Al(p,$\gamma$)$^{28}$Si  	&	 [HA00]$^{a}$,[I]$^{a}$ 	&	  	\\
$^{14}$N($\alpha$,n)$^{17}$F  	&	  	&	  	&	$^{27}$Al(p,$\alpha)^{24}$Mg  	&	 [I]$^{a}$ 	&	  	\\
$^{15}$N(p,$\gamma$)$^{16}$O  	&	  	&	  	&	$^{27}$Al($\alpha$,n)$^{30}$P  	&	  	&	  	\\
$^{15}$N(p,n)$^{15}$O  	&	  	&	  	&	$^{28}$Si(p,$\gamma$)$^{29}$P  	&	 [I]$^{a}$ 	&	  	\\
\hline

\end{tabular}

\end{table*}
 
NACRE is described in detail by Angulo et al. (1999). It contains in
particular

\noindent (1) the formalism that has been adopted in order to derive the
Maxwellian-averaged astrophysical rates of the (exothermic and endothermic) charged particle
induced reactions and of their reverse;  
 
\noindent (2) a general description of the treatment of the data, which has sometimes to be
adapted to specific cases, as described in the comments accompanying each of the compiled reactions.
If necessary, tables of narrow resonances with their characteristics are provided.  A careful analysis
of the experimental uncertainties in the quantities involved in the evaluation of the reaction rates is
carried out for each reaction. From this, {\it adopted} rates are provided, as well as {\it low} and
{\it high} limits. This large-scale estimate of the uncertainties is considered to be an essential
feature of the NACRE compilation;

\noindent (3) the procedure adopted for extrapolating experimental data when required in order to
evaluate the reaction rates. This extrapolation to the very low energies necessary
to evaluate the rates down to the lowest considered temperatures raises many difficult
problems. A particular one concerns the necessity to correct the laboratory cross sections
for electron screening (e.g. Rolfs \& Rodney 1988) before their use in the rate calculations.
This effect becomes significant for relative energies $E$ of the reaction partners such that
$E/U_{\rm e}\leq 100$, where $U_{\rm e}$ is the so-called screening  potential. An
approximate procedure is devised in order to eliminate data that can be `polluted' by
laboratory screening. On the other hand, in stellar conditions, the nuclei are
surrounded by a dense electron gas that reduces the Coulomb repulsion and makes
the penetration of the Coulomb barrier easier. The cross sections are therefore enhanced in
comparison with those of reactions between bare nuclei. This stellar screening effect can
be evaluated by applying, for example, the Debye-H\"{u}ckel theory (e.g. Cox \& Giuli 1968).
The NACRE rates {\it exclude} these stellar screening factors. 
  
The evaluation of the rates for temperatures as high as $T_9 = 10$ also requires the
application of special extrapolation techniques when reliable cross section data are
lacking at sufficiently high energies. In those cases, a duly documented HF
approach is used (see also Sect. \ref{th-rates-general}), which smoothly connects to the
experimentally-based rate estimates;

\noindent (3) the procedure used to evaluate the contribution of excited states of the
target nuclei to the effective stellar reaction rates. In a stellar plasma, the excited
levels of a target nucleus are indeed thermally populated, and thus contribute to the
reaction mechanism. As a result, the stellar rates may differ from those obtained when the
target nuclei are in their ground state. This difference is expressed in terms of a
correction factor $r_{tt}$ that has to multiply the target ground state rate in order to
obtain the stellar rate. In general, this correction cannot be derived experimentally. 
The very approximate treatment of this correction by Caughlan
\& Fowler (1988) is replaced in NACRE by a more quantitative procedure based on the use of
the HF model (Sect. \ref{th-rates-general}), and on the classical assumption of a
Maxwellian population of the nuclear excited states. Note that this assumption may be
invalid if the target nucleus has an isomeric state which is not in thermal equilibrium with the 
ground state in certain astrophysical situations. Such an example of astrophysical
interest concerns \chem{26}{Al}, to which a special treatment is applied.                   
 
The calculation of $r_{tt}$ requires in particular the knowledge of the
temperature-dependent partition functions of the target nuclei normalized to their ground
state 

\begin{eqnarray}                                                                         
G_i(T) = \frac{1}{2J^0_i+1} \sum_\mu (2J^\mu_i
+1)\,                                                                          
\exp \left( -\frac {\epsilon^\mu_i}{kT}
\right),                                                                         
\label{partf}                                                                         
\end{eqnarray}                                                                         
                                                                         
\noindent where the summation extends over the states $\mu$ of target $i$ with excitation
energy $\epsilon^\mu_i$ and spin-parity $J^\mu_i$ ($\mu = 0$ for the ground state);

\noindent (4)  The values of the adopted, low and high Maxwellian-averaged unscreened rates {\it on the
ground state target} are provided in tabular form for a selection of temperatures in the $0.001 \le T_9
\le 10$ range ($T_9$ is the temperature in billion K). As they may be desired by some users, analytical
formulae approximating the tabulated {\it adopted} ground state rates or those including the
contribution of thermalized excited states are provided as well. It has to be noted that some of
these expressions have a form deviating from those classically advertised in the nuclear
astrophysics literature (e.g. Caughlan \& Fowler 1988; Rolfs \& Rodney 1988; see e.g.
Eq.~(23) of Angulo et al. 1999). They are considered to provide more secure approximations
to the numerically calculated Maxwellian-averaged rates. In addition, NACRE
provides analytical formulae for the multiplicative factor that has to be applied to the
rate of a given reaction in order to evaluate the rate of the inverse process, as well as
for the nuclear partition functions for the targets involved in the compiled reactions (see
Eq.~\ref{partf}).  

NACRE is accessible electronically through the BRUSLIB library website {\it
http://www-astro.ulb.ac.be}. Much more material is available there than presented above or
published by Angulo et al. (1999). In particular, reaction rates are available for far more
extended temperature grids. In addition, graphical material (astrophysical
$S$-factors) can be retrieved. 

Finally, it is noted that other astrophysics-oriented experimentally-based reaction rate
compilations have appeared around the time of the NACRE publication (Adelberger et al.
1998), or later (Iliadis et al. 2001; Descouvemont et al. 2004). The reactions considered in the latter
two compilations are identified in Table~1, which also provides references to additional data or
evaluations of relevance. This additional information will of course be duly analyzed and integrated
into the new version of NACRE currently in preparation.

\section{The BRUSLIB theoretical reaction rates}
\label{th-rates}

Much effort has been devoted in the last decades to measurements of 
reaction cross sections for astrophysical purposes. As already stated in
Sect.~\ref{intro}, difficulties related to the conditions prevailing in the
astrophysical plasmas remain, however. In particular, charged-particle induced reactions at
stellar energies (far below the Coulomb barrier) have extremely small cross sections that are
highly difficult to measure. Specific difficulties are also raised by the measurements of
photoreactions that have to be included in many astrophysics models (e.g. Arnould \& Goriely
2003 for a review). In addition, a huge number (thousands) of reactions of relevance involve
more or less exotic nuclides. Clearly, many of these reaction rate experiments will remain
unfeasible for a long time to come. Theory has thus to supply the necessary data, which
represents a major challenge of its own.

BRUSLIB provides astrophysicists with
a very extended set of thermonuclear rates in the $0.01 \le T_9 \le 10$
temperature range for all the
reactions induced by neutron, proton and
$\alpha$-particle captures by all nuclei with $Z,N \ge 8$ and $Z<110$  located between the neutron and proton drip lines
 (i.e some 8000 nuclei). The rates of the ($\gamma$,n), ($\gamma$,p) and
($\gamma$,$\alpha$) photodisintegrations of all these nuclides are also tabulated for the same
temperature grid. The calculations rely on the code MOST (Goriely 1998; Arnould \& Goriely 2003)
based on the HF model (Hauser \& Feshbach 1952). This
approach is valid if the density of nuclear levels of the compound systems formed as a
result of these captures is high enough. This is indeed  the case if the targets are
heavy enough and if they are located far enough from the proton or neutron drip lines to
ensure that the excitation energies of the compound systems are high enough. 
Roughly speaking, the reliability of the BRUSLIB rates may thus be limited to nuclides with mass
numbers $A \gsimeq 40$ close enough to the valley of stability, this mass limit being of course shifted
to higher and higher values as one moves further and further away from the valley.
If these constraints are not met, a non-statistical treatment is more
suited (Goriely 1997). It has to be noticed that purely theoretical HF rates are provided for some
reactions already included in NACRE (Table~1). In these cases, it is advised to adopt the NACRE rates,
 or those from the Iliadis et al. (2001) or Descouvemont et al. (2004) compilations, as these
rely on experimental data.
  
A few words are in order at this point to clarify the basic philosophy governing the
selection of the models necessary for the calculation of the various ingredients
entering the HF BRUSLIB rates. Their evaluation is based on {\it global} and {\it
coherent} {\it microscopic} (or at least semi-microscopic) models. These 
features have an importance in the astrophysics modellings that cannot be underestimated,
and make the BRUSLIB rate library unique. They are indeed not shared by HF calculations
based on other codes, like Talys (Koning et al. 2002), Empire (Herman et al. 2002), or
Non-Smoker (Rauscher \& Thielemann 2001). 

The global character of the underlying models is made highly desirable by the fact that, for
many specific applications including nuclear astrophysics, a very large body of reaction
rates for which no experimental data exist have to be provided. 
The microscopic nature of the underlying models is essential as well. For nuclear
astrophysics, as well as in other fields, a large amount of data need to be
extrapolated far away from experimentally known regions. In these situations, two major
features of the nuclear theories have to be considered. The first one is the {\it accuracy}
of a model. In most nuclear applications, this criterion has been the main, if not the
unique, one for selecting a model. The second one is the {\it reliability} of the
predictions. A physically sound model that is based on first principles and is as close as
possible to a microscopic description of the nuclear systems is expected to provide the
best possible reliability of extrapolations far away from experimentally known regions.
Of course, the {\it accuracy} of such microscopic models in reproducing experimental
data may be poorer than the one obtained from more phenomenological models in which
enough free parameters can guarantee a satisfactory reproduction of the data at the
expense of the quality of the input physics, and consequently of the reliability. The
coherence (or `universality') of these microscopic models (through e.g. the use of the
same basic nuclear inputs, like the effective nuclear forces) is also required as
different ingredients have to be predicted in order to evaluate each reaction rate.
Failure to meet this requirement could lead to bad ways in the rate evaluations.

The BRUSLIB MOST rates are based on microscopic (or at least
semi-microscopic) models for the necessary nuclear structure ingredients or interaction
properties (Sect.~\ref{th-rates-general}) that have been constructed to be global and
coherent  to the largest possible extent. In addition, these models have the major advantage
of reaching a satisfactory compromise between accuracy and reliability for the relevant
nuclear inputs, and consequently for the rates themselves. The level of accuracy of these
models is in fact fully comparable to the one obtained from available more or less highly
parametrized phenomenological approaches, while their reliability is by far better.

Finally, the BRUSLIB reaction rates duly take into account the necessary astrophysical
specificities. The temperature dependence of the rates is predicted from
the consideration of the Maxwell-Boltzmann distribution of the relative velocities of the
reaction partners and of the target nuclear excited states states (par. (3) of
Sect.~\ref{NACRE}). In contrast, the provided rates are not corrected for stellar electron
screening effects.
 
\subsection{Reaction rate calculations: general framework}
\label{th-rates-general}

Some basics of the HF formalism adopted in the code MOST  
are briefly recalled in e.g. Arnould \& Goriely (2003), and are not repeated here. The limits of its
validity are also reminded above.  Let us just point out the following:
 
\noindent (1) under local thermodynamic equilibrium conditions,  
the effective stellar rate of  $I + j\rightarrow L + k$ per pair of particles in the 
entrance channel at temperature $T$ taking due account of the contributions of the various excited
states $\mu$ of the target is expressed in a classical notation as
%
\begin{eqnarray}
&& N_A\langle\sigma v\rangle^*_{jk}(T) =  \Bigl( \frac{8}{\pi m} \Bigr) ^{1/2}
                  \frac{N_{A}}{(kT)^{3/2}~G(T)} \times \cr
       & &    \quad  \int_{0}^{\infty} \sum_{\mu} \frac{(2J_I^{\mu}+1)}{(2J_I^{0}+1)}
                  \sigma_{jk}^{\mu}(E)E\exp \Bigl(-\frac{E+\varepsilon^{\mu}_I}{kT}\Bigr) dE,
\label{eq_rate}
\end{eqnarray}
%
where $k$ is the Boltzmann constant,
$m$ the reduced mass of the $I^0 + j$ system, $N_{A}$ the Avogadro number, 
$\sigma^\mu_{jk}(E)$ the cross section at relative energy $E$ of the $I^{\mu} + j\rightarrow L + k$
reaction, and
$G(T)$ is the partition function given by Eq.~\ref{partf}, where  $J_I^\mu$ and
$\epsilon_I^\mu$ are defined. The BRUSLIB photodisintegration
rates are estimated by applying the reciprocity theorem on the radiative capture rates derived
from Eq.~\ref{eq_rate}. This procedure leads to
%
\begin{eqnarray}
\lambda_{(\gamma,j)}^*(T) &=& \frac{(2J_I+1)(2J_j+1)}{(2J_L+1)}~\frac{G_I(T)}{G_L(T)}
\Bigl( \frac{A_I A_j}{A_L}\Bigr)^{3/2} \times \cr
& & \Bigl( \frac{kT}{2 \pi \hbar^2 N_A} \Bigr)^{3/2} N_A \langle\sigma v\rangle^*_{(j,\gamma)}
~ {\rm e}^{-Q_{j\gamma} /kT},
\label{eq_inverserate}
\end{eqnarray}
%
where $Q_{j\gamma}$ is the Q-value of the $I^0(j,\gamma)L^0$ capture. Note that,
in stellar conditions, the reaction rates for targets in thermal equilibrium 
obey reciprocity since the forward and reverse channels are symmetrical, in contrast to the
situation which would be encountered for targets in their ground states only (Holmes et
al. 1976).  

The uncertainties involved in any HF prediction are dominated by those involved in the
evaluation of the nuclear quantities necessary for the calculation of the cross sections, such as
the masses, deformations, matter distributions, single-particle levels, and level densities of
target and residual nuclei, as well as the optical potentials. Special problems are also raised by
the evaluation of the photon widths $T_\gamma$. The interested reader is referred to e.g. Arnould
\& Goriely (2003) for more details.

\subsection{Nuclear masses, level densities, and partition functions}
\label{masses}

The BRUSLIB MOST predictions rely on the experimental nuclear mass data compiled by Audi
et al. (2003). When not measured (a very common situation for various astrophysics
applications), use is made of the HFB-9 microscopic mass model (Goriely et al. 2005).
This model also provides the necessary information on nuclear deformation, charge and
matter distributions, pairing properties and single-particle spectra. The nuclear level
densities are extracted from a microscopic model developed by Goriely (1996) (see also
e.g. Demetriou \& Goriely 2001).

The nuclear partition functions $G_i(T)$ entering the evaluation of the
astrophysical reaction rates (Sects.~\ref{NACRE} and \ref{th-rates}) are calculated from
(i) a summation over experimentally known levels (Eq.~\ref{partf}) up to an excitation
energy $\epsilon^\omega$ above which the knowledge of the energy spectrum is considered
to be incomplete, and (ii) a generalization of
Eq.~\ref{partf} involving an integration over a level density evaluated as described
above.
 
\subsection{Optical potentials}
\label{th-pot}

Phenomenological optical potentials (OPs) (generally of the Woods-Saxon type) may not be
well suited for certain applications, particularly those involving exotic nuclei.  It is
considered profitable to use more microscopically-based potentials, whenever possible. A
semi-microscopic OP, usually referred to as the JLM potential (Jeukenne et al. 1977), is
available for the description of the nucleon-nucleus case. This OP has been revised
recently for nucleons incident on spherical or quasi-spherical nuclei with masses
$40\le A\le 209$ at energies ranging from 1 keV to 200 MeV (Bauge et al. 2001). The
resulting new version gives a global
satisfactory agreement with experimental data, even if some
improvements would be most welcome, especially in the low-energy domain and in the
treatment of deformed or exotic nuclei. It is adopted for the BRUSLIB rate
evaluations. 

The situation for the $\alpha$-particle-nucleus OPs is much less satisfactory, and one
still has to rely on phenomenological potentials.  Most of the proposed OPs are derived
from fits to elastic $\alpha$-nucleus scattering data at energies $E \gsimeq 80~{\rm
MeV}$ or, in some cases, to (n,$\alpha$) cross sections at lower energies. However, the
OP, and in particular its imaginary component, is known to depend strongly on energy
below the Coulomb barrier. As a consequence, its extrapolation to sub-Coulomb energies
that is necessitated by astrophysics applications is more insecure than in the case of
nucleons. Several attempts to device  a global
$\alpha$-nucleus OP for the description of the scattering and reaction cross sections at
energies $E \lsimeq 20$ MeV of better relevance to astrophysics have been conducted (e.g.
Arnould \& Goriely 2003 for more details and references). The scarcity of  experimental
data, particularly in the $A > 100$ mass range, limits dramatically the predictive power
of any of the constructed global OPs. This has the immediate consequence of reducing the
reliability of the rate predictions as they depend sensitively on the
$\alpha$-particle-nucleus OPs.  The BRUSLIB $\alpha$-capture rates are calculated
with the global OP III developed by Demetriou et al. (2002).
 
\subsection{$\gamma$-ray strength function}
\label{th-strength}

When applied to radiative captures, the total photon transmission coefficient entering the
calculation of the $\sigma^\mu_{j\gamma}$ cross section of Eq.~\ref{eq_rate} is dominated by the
$E1$ transitions. The calculation of the
$E1$-strength function necessitates the knowledge of the low-energy tail of the Giant
Dipole Resonance (GDR) of the compound system formed in the reaction process. The photon
transmission coefficient is  most frequently described in the framework of the
phenomenological generalized Lorentzian model (e.g. Goriely 1998). 
%
%
%
%
The Lorentzian GDR approach suffers, however, from shortcomings of various sorts. On the one
hand, it is unable to predict the enhancement of the $E1$ strength at energies below the
neutron separation energy demonstrated by nuclear resonance fluorescence experiments. This
departure from a Lorentzian profile may manifest itself more clearly for neutron-rich nuclei,
and especially in the form of a so-called pygmy $E1$ resonance (Govaert et al. 1998,
Zilges et al. 2002). On the other hand, even if a Lorentzian function provides a suitable
representation of the
$E1$ strength, the location of its maximum and its width remain to be predicted from some
underlying model for each nucleus. For astrophysical applications, these properties have
often been obtained from a droplet-type of model (Myers et al. 1977), which clearly
lacks reliability when dealing with exotic nuclei. This introduces large uncertainties in
certain rate estimates (e.g. Goriely \& Khan 2002, Goriely et al. 2004)

In view of this situation, it is clearly of substantial interest to develop models of the
microscopic type which are hoped to provide a reasonable reliability and predictive
power for the $E1$-strength function. Various attempts of this sort have been conducted
(e.g. Arnould \& Goriely 2003, and references therein). The BRUSLIB MOST rate
calculations are based on the semi-microscopic QRPA $E1$ model developed by Goriely et
al. (2004).

\section{A confrontation between measured and calculated reaction rates}
\label{th-exp}

In order to evaluate the overall quality of the BRUSLIB reaction rate predictions, this
section presents a confrontation between selected experimental data and calculations in
which only the ground state contribution ($\mu = 0$) is taken into account in
Eqs.~\ref{eq_rate} and \ref{eq_inverserate}, the consideration of target excited states being
irrelevant in the laboratory conditions.  
 
\begin{figure}[h]
 \begin{center}
\includegraphics[scale=0.30]{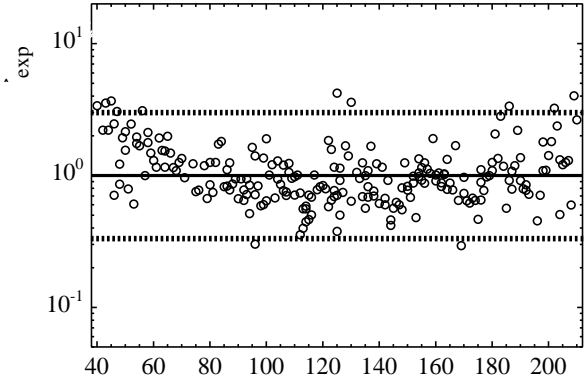}
\caption[]{Comparison of BRUSLIB Maxwellian-averaged (n,$\gamma$) rates $\langle\sigma v
\rangle_{\rm th}$ with experimental values (Bao et al. 2000) at $T=3.5 \times
10^8{\rm K}$}
\label{fig_ng_exp}
\end{center}
\end{figure}

\begin{figure}
\includegraphics[scale=0.40]{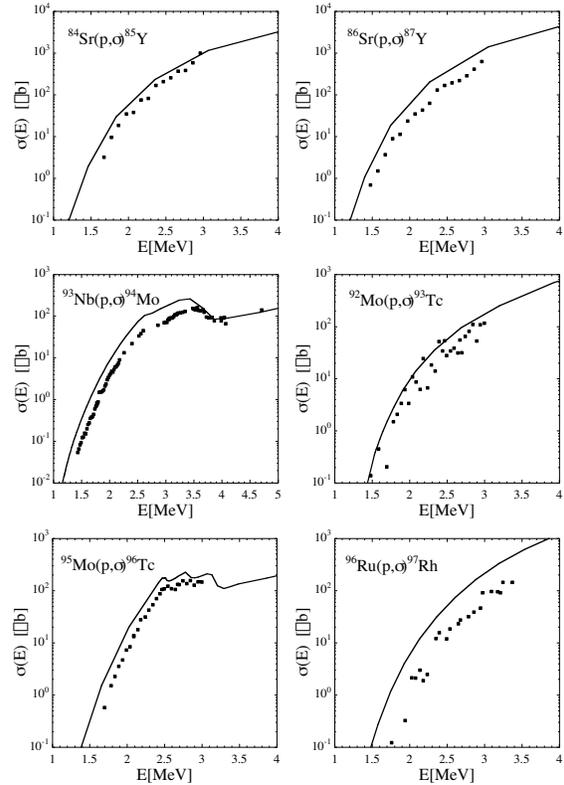}
\caption[]{Comparison between the BRUSLIB predictions and the measured cross sections of some
(p,$\gamma$) reactions on targets heavier than \chem{84}{Sr} (see Arnould \& Goriely 2003 for
references to the laboratory work)}
\label{fig_pg_exp}
\end{figure}

Figure~\ref{fig_ng_exp} compares the BRUSLIB predictions of the Maxwellian-averaged
(n,$\gamma$) rates $\langle\sigma v\rangle$ at
$T = 3.5 \times 10^8$ K with experimental data for some 228 nuclei heavier than
$^{40}{\rm Ca}$ included in the compilation of Bao et al. (2000). It appears that the
calculations agree with all data to within a factor of three. Figure~\ref{fig_pg_exp} compares
the experimental cross sections for some low-energy (p,$\gamma$) reactions on targets
heavier than Fe with the corresponding BRUSLIB rates. Again, and broadly speaking,
the agreement is very satisfactory for the vast majority of the (p,$\gamma$) data. It is
difficult, however, to be much more specific, as the quality of this agreement may depend more or less 
drastically on temperature, in contrast to the situation encountered in the neutron capture case. In
addition, the comparison is limited to targets up to Sn. Experiments with heavier nuclei would be most
welcome, as they might help refining the  predictions. 
 
\begin{figure}
\includegraphics[scale=0.25]{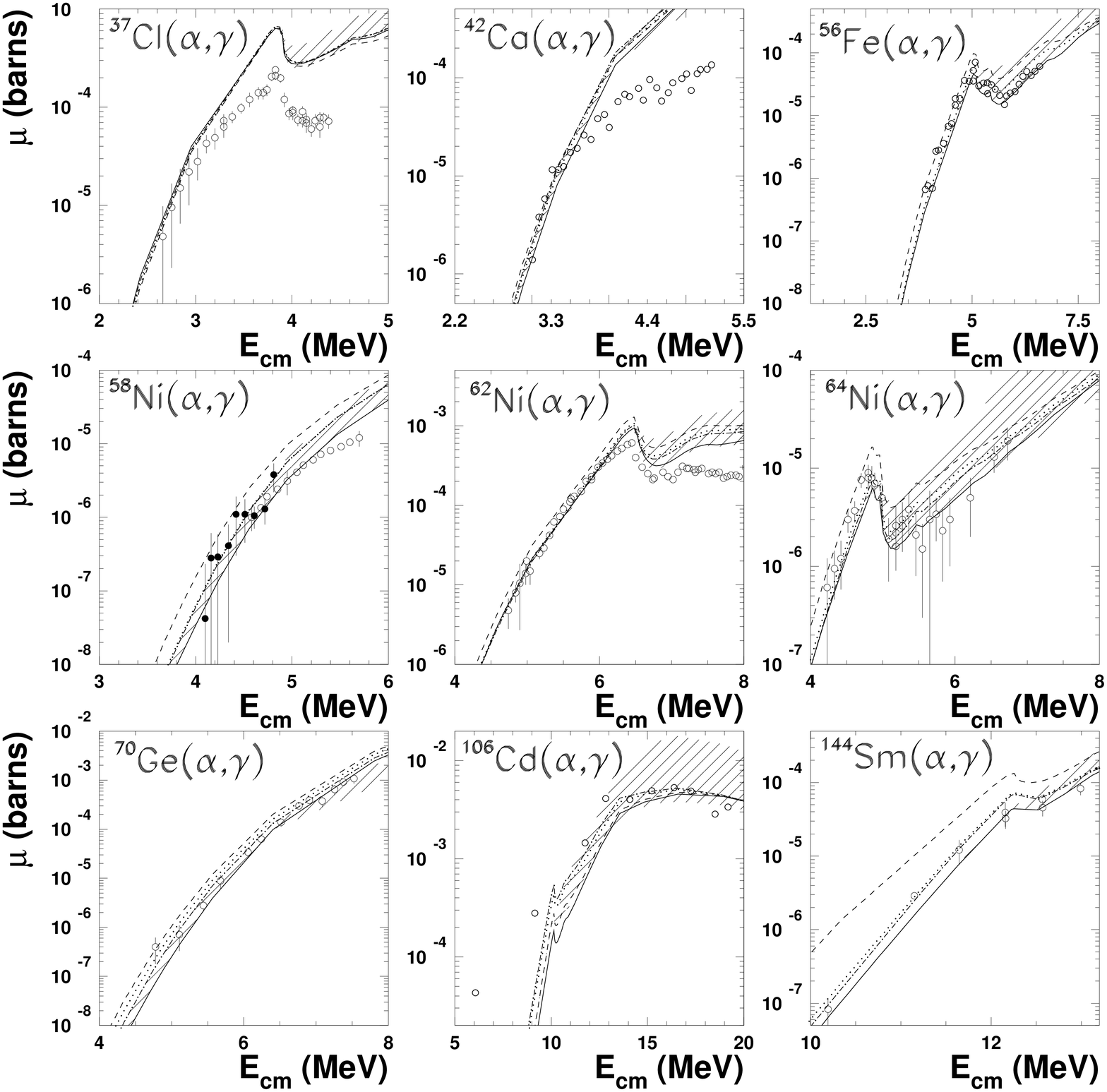}
\caption{Cross sections for the ($\alpha$,$\gamma$) reactions on $^{37}$Cl, $^{42}$Ca,
$^{56}$Fe, $^{58}$Ni, $^{62,64}$Ni, $^{70}$Ge, $^{106}$Cd and
$^{144}$Sm (see Arnould \& Goriely 2003 for references to the laboratory work). Black and
open circles represent experimental data. The solid lines give the calculated cross sections  
used to construct the BRUSLIB rates}
\label{fig_ag}
\end{figure}
 
The situation is by far less clear for the ($\alpha$,$\gamma$) reactions. This results from
the lack of a large enough body of experimental data for sub-Coulomb cross sections
combined with the difficulties to construct global and reliable
$\alpha$-nucleus OPs (Sect.~\ref{th-pot}). These theoretical problems
are magnified by the fact that, at the sub-Coulomb energies of astrophysical relevance,
the reaction rate predictions are highly sensitive to these potentials through the
corresponding $\alpha$-particle transmission coefficients. Figure~\ref{fig_ag} displays a
comparison between some low-energy measurements of ($\alpha$,$\gamma$) cross sections and
MOST predictions used for evaluating the BRUSLIB rates. The quality of the agreement appears
to vary from case to case, and is also highly sensitive to temperature. Uncertainties in the
calculated values originate mainly from those in the nuclear level densities and
$\alpha$-nucleus OPs (see Arnould \& Goriely 2003 for details, and
in particular for a discussion of the \reac{144}{Sm}{\alpha}{\gamma}{148}{Gd} reaction of
special astrophysics interest). 

\section{The nuclear network generator NETGEN}
\label{NETGEN}
The {\it Nuclear Network Generator} NETGEN is an interactive Web-based tool to generate
Maxwellian-averaged nuclear reaction rates for networks and temperature grids specified by the
user. It is fully documented at the web address {\it http://www-astro.ulb.ac.be/Netgen}. 

NETGEN relies mainly on the BRUSLIB/NACRE library. It also makes use of 

\noindent (1) the post-NACRE compilations by Iliadis et al. (2001) and Descouvemont et al. (2004) (see
Table~1); 

\noindent (2) experimentally-based published rates for about 70 charged-particle induced
reactions not included in NACRE. Specific references are provided for each of these reactions;

\noindent (2) more than 200 experimentally-based radiative neutron capture rates. Most of these are
adopted from Bao et al. (2000). Specific references are given in each case;
 
\noindent (3) $\beta$-decay and electron-capture rates, including (i) laboratory measurements
compiled by Horiguchi et al. (1996), (ii) theoretical estimates based on the evaluation of
individual transitions (Takahashi \& Yokoi 1987), on the gross theory (Tachibana et al. 1990), or
on the more microscopic ETFSI + cQRPA model (Borzov \& Goriely 2000).

 For the sake of completeness, the references of Table~1 that are not contained in the NACRE,
Iliadis et al. (2001) or Descouvemont et al. (2004) compilations are also included in NETGEN.
 
For each reaction or $\beta$-decay rate, NETGEN selects by default the data source that
is considered to be the most reliable. We select in order of preference  the latest available
compilation (NACRE, Iliadis et al. 2001, or Descouvemont et al. 2004), experimental data, detailed
microscopic calculations, and BRUSLIB rates derived from global calculations. The user may nevertheless
adopt another choice for selected cases by specifying `bibliographic indexes' for each reaction, the
table of rates being accompanied by a `log file' listing the selected data source among all those
available in the library. A FORTRAN program handling the rates is also made available. Note that all
rates duly take into account the contribution from the excited states of the target nuclei, as
discussed in the previous sections.

Various NETGEN options are currently offered through the web interface, as described in detail at
{\it http://www-astro.ulb.ac.be/Netgen}:

\noindent (1) Generate a table of reaction rates on a temperature grid for a network that has been
\begin{itemize}
\item typed in reaction by reaction (offering the possibility to select non-default rates, as
mentioned above), or  
\item generated automatically between interactively-selected  boundaries on the proton 
    and mass numbers, and involving various possible sets of reactions (p-,
n-, $\alpha$-captures, $\beta$-decays and/or photodisintegrations), or
\item uploaded on the server.
\end{itemize} 
\noindent (2) Plot individual reaction rates, and provide .gif or .ps files. 
 
\section{Conclusions}
\label{concl}

This is the first release in an astronomy and astrophysics journal of the BRUSLIB nuclear
reaction rate library and of the nuclear network generator NETGEN. The format of the packages is
chosen in order to make them easily accessible to astrophysicists and to be well suited for a large
variety of astrophysics needs. They are made available through the web site
{\it http://www-astro.ulb.ac.be}.

The BRUSLIB NACRE package contains a detailed experimentally-based evaluation and compilation of
the rates of 86 proton or $\alpha$-capture reactions on (mainly) stable targets up to Si for
temperatures ranging from $10^6$ to $10^{10}$ K. The electronic files contain much more information
than published by Angulo et al. (1999).  

The NACRE data are complemented with about 100000 thermonuclear rates of nucleon and
$\alpha$-captures on about 8000 $8 \leq Z \leq 110$ nuclei located between the proton and neutron
drip lines. The calculations are based on a statistical Hauser-Feshbach model featuring a
microscopic (or at least very close to microscopic) evaluation of the basic ingredients of the
model. These predictions are seen to compare favourably with the limited set of experimental
reaction cross section data on intermediate-mass and heavy nuclei at energies close to those
of astrophysics relevance. The rates of photodisintegration of the whole set of nuclei are also
provided. They are derived from the application of the reciprocity theorem.

NETGEN is an interactive web-based tool allowing the construction on a user-friendly basis of
nuclear reaction networks specified by the user on a temperature grid of his/her choice. A full
documentation of its use can be found at the web address {\it http://www-astro.ulb.ac.be/Netgen}.

It is hoped that the easy availability of a very large set of nuclear reaction rate evaluations and
predictions, as well as of a nuclear network generator will be helful to many researchers for a
large variety of applications. Quite clearly, this enterprise is of a highly dynamical and
long-term nature. BRUSLIB will be continuously improved and expanded, and new
releases will be made accessible to the community every time a substantial enough body of new data
becomes available.

\begin{acknowledgements}
The authors thank all the collaborators who have made possible the development over the years of the
 Brussels library of nuclear data, and in particular M. Pearson, P. Demetriou and E. Khan.  This work
has been supported in part by the Interuniversity Attraction Pole IAP 5/07 of the Belgian Federal
Science Policy and by the Konan University - Universit\'e Libre de Bruxelles convention `Construction of
an Extended Nuclear Database for Astrophysics'. S.G. is FNRS Research Associate. A.J. is FNRS Senior
Research Associate. 
\end{acknowledgements}

%

\end{document}